# High Speed Reconfigurable FFT Design by Vedic Mathematics

Ashish Raman, Anvesh Kumar and R.K.Sarin

**Abstract**—The Fast Fourier Transform (FFT) is a computationally intensive digital signal processing (DSP) function widely used in applications such as imaging, software-defined radio, wireless communication, instrumentation. In this paper, a reconfigurable FFT design using Vedic multiplier with high speed and small area is presented. Urdhava Triyakbhyam algorithm of ancient Indian Vedic Mathematics is utilized to improve its efficiency. In the proposed architecture, the 4x4 bit multiplication operation is fragmented reconfigurable FFT modules. The 4x4 multiplication modules are implemented using small 2x2bit multipliers. Reconfigurability at run time is provided for attaining power saving. The reconfigurable FFT has been designed, optimized and implemented on an FPGA based system. This reconfigurable FFT is having the high speed and small area as compared to the conventional FFT.

**Index Terms**-Digital Signal Processing, FPGA, Urdhava Triyakbhyam, FFT

——————————  ◆  ——————————

## 1. INTRODUCTION

Digital signal processing (DSP) is the technology that is omnipresent in almost every engineering discipline. Faster additions and multiplications are of extreme importance in DSP for convolution, discrete Fourier transform, digital filters, etc. The core computing process is always a multiplication routine; therefore, DSP engineers are constantly looking for new algorithms and hardware to implement them. Vedic mathematics is the name given to the ancient system of mathematics, which was rediscovered, from the Vedas between 1911 and 1918 by Sri Bharati Krishna Tirthaji. The whole of Vedic mathematics is based on 16 *sutras* (word formulae) and manifests a unified structure of mathematics. As such, the methods are complementary, direct and easy.

————————————————

- A.Raman is with National Institute of Technology, Jalandhar
- Anvesh kumar is with National Institute of Technology, Jalandhar
- R.K.Sarin is with National Institute of Technology, Jalandhar

Due to a growing demand for such complex DSP application, high speed, low cost system-on-a-chip (SOC) implementation of DSP algorithm are receiving increased the attention among the researchers and design engineer. Fast Fourier Transform (FFT) is the one of the fundamental operations that is typically performed in any DSP system. Basic formula of computation of FFT is

$$X(k) = \sum_{n=0}^{N-1} x(n) \, W_N^{nk} \quad 0 \leq k \leq N-1$$

The Fast Fourier Transform (FFT) is a computationally intensive digital signal processing (DSP) function widely used in applications such as imaging, software-defined radio, wireless communication, instrumentation and machine inspection. Historically, this has been a relatively difficult function to implement optimally in hardware leading many software designers to use digital signal processors in soft implementations.

Vedic Mathematics is the ancient system of mathematics which has a unique technique of calculations based on 16 Sutras. Employing these techniques in the computation algorithms of the coprocessor will reduce the complexity, execution time, area, power etc. In this paper reconfigurable FFT is proposed to design by Vedic mathematics. Urdhva tiryakbhyam, being a general multiplication formula, is equally applicable to all cases of multiplication. Nikhilam algorithm with the compatibility to different data types. This sutra is to be used to build a high speed power efficient reconfigurable FFT.

Paper is organized as follow: section II describes the integration of two fields i.e., FPGA technology and DSP and discusses the role FPGA has played in implementing DSP algorithm. Section III describe the basic and introduction to the Vedic algorithm and how this algorithm are used to solve the different problem of mathematics and this chapter also include the review of literature survey. Section IV describes the basic introduction of the FFT and how the different modules of the FFT are designed by Vedic algorithm. Section V describes the various results.

## 2. INTEGRATION OF DSP WITH FPGA

In electronic system design, the main attraction of microprocessors/microcontrollers is that it considerably lessens the risk of system development by reducing design complexity. As the hardware is fixed, all of the design effort can be concentrated on developing the *code* which will make the hardware work to the required system specification. This situation has been complemented by the development of efficient software compilers which have largely removed the need for designer to create assembly language; to some extent, this can absolve the designer from having a detailed knowledge of the microprocessor architecture (although many practitioners would argue that this is essential to produce *good* code). This concept has grown in popularity and embedded microprocessor courses are now essential parts of any electrical/electronic or computer engineering degree course.

A lot of this process has been down to the software developer's ability to exploit underlying processor architecture, the Von Neumann architecture. However, this advantage has also been the limiting factor in its application to the topic of this text, namely digital signal processing (DSP). In the Von Neumann architecture, operations are processed sequentially, which allows relative Straight forward interpretation of the hardware for programming purposes; however, this severely limits the performance in DSP applications which exhibit typically, high levels of parallelism and in which, the operations are highly data independent – allowing for optimizations to be applied.

This limitation is overcome in FPGAs as they allow what can be considered to be a second level of programmability, namely programming of the underlying processor architecture. By creating architecture that best meets the algorithmic requirements, high levels of performance in terms of area, speed and power can be achieved. In high volumes, ASIC implementations have resulted in the most cost effective, fastest and lowest energy solutions. However, increasing mask costs and impact of 'right first time' system realization have made the FPGA, a much more attractive alternative. In this sense, FPGAs capture the performance aspects offered by ASIC implementation, but with the advantage of programmability usually associated with programmable processors. Thus, FPGA solutions have emerged which currently offer several hundreds of giga operations per second (GOPS) on a single FPGA for some DSP applications which is at least an order of magnitude better performance than microprocessors.

## 3. VEDIC ALGORITHM

The proposed Vedic multiplier is based on the Vedic multiplication formulae (Sutras). These Sutras have been traditionally used for the multiplication of two numbers in the decimal number system. In this work, we apply the same ideas to the binary number system to make the proposed algorithm compatible with the digital hardware. Vedic multiplication based on some algorithms, some are discussed below:

### 3.1 URDHVA TIRYAKBHYAM SUTRA

Urdhva Tiryakbhyam (Vertically and Crosswise), deals with the multiplication of numbers. This Sutra has been traditionally used for the multiplication of two numbers in the decimal number system. In this project, we apply the same idea to the binary number system to make it compatible with the digital hardware. Let us first illustrate this Sutra with the help of an example in which two decimal numbers are multiplied. Line diagram for the multiplication of two numbers (234× 316) is shown in Fig. 1. The digits on the two ends of the line are multiplied and the result is added with the previous carry. When there are more lines in one step, all the results are added to the previous carry. The least significant digit of the number thus obtained acts as one of the result digits and the rest act as the carry for the next step. Initially the carry is taken to be zero.

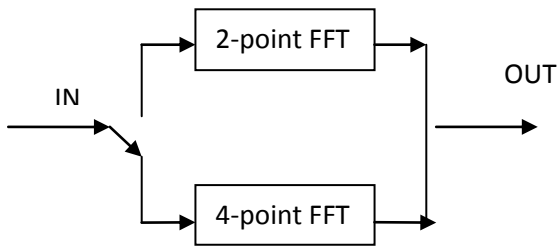

Fig. 1. Multiplication of two decimal numbers by Urdhva Tiryakbhyam Sutra.

## 4. RECONFIGURABLE ARCHITECTURE

The basic architecture of the reconfigurable FFT is to be shown in Fig.2. It can calculate Fourier transform either 2-point or 4-point depending upon the select line which is to be given by the programmer. Its calculate 2-point and 4-point FFT without changing the hardware of the system. This particular reconfigurable FFT is to be designed by using Vedic adder, Vedic sub tractor, and Vedic multiplier. The delay produced by the Vedic reconfigurable FFT is smaller than the delay produced by the conventional reconfigurable FFT.

Fig.2 Reconfigurable architecture

## 5. RESULT AND DISCUSSION:

From this table it's clear that how the Vedic reconfigurable FFT having less delay and small area as compare to the reconfigurable FFT. Fig.5. show the simulation result of Vedic reconfigurable FFT. The Vedic reconfigurable FFT is implemented into the Vertex 2 pro, Device-XC2VP2, package - FG256, speed:-6.

Table. 1 delay comparision

| Architecture | Conventional FFT | Vedic FFT |
|---|---|---|
| 2-point | 8.532ns | 8.251ns |
| 4-point | 12.543ns | 11.947ns |

Table. 2 delay comparision for reconfigurable architecture

| Architecture | Delay |
|---|---|
| Reconfigurable FFT | 13.931 ns |
| Vedic Reconfigurable FFT | 13.325 ns |

Table 3. area comparision

| Type of the Architecture | Area used | |
|---|---|---|
| | Number of Slices: | Number of 4 input LUTs: |
| Reconfigurable FFT | 81 out of 1408   5% | 147 out of 2816   5% |
| Vedic Reconfigurable FFT | 69 out of 1408   4% | 127 out of 2816   4% |

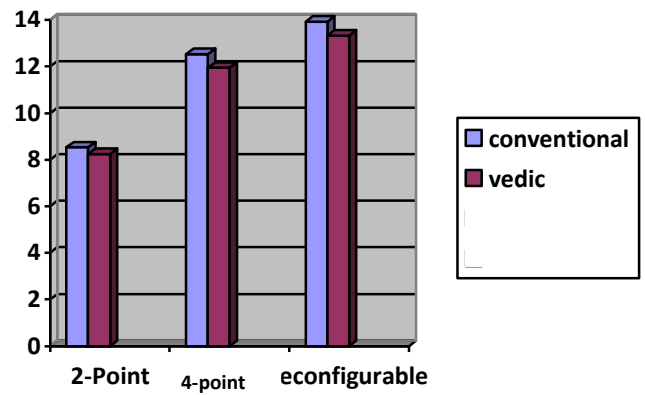

Fig3. Graph of delay comparisons

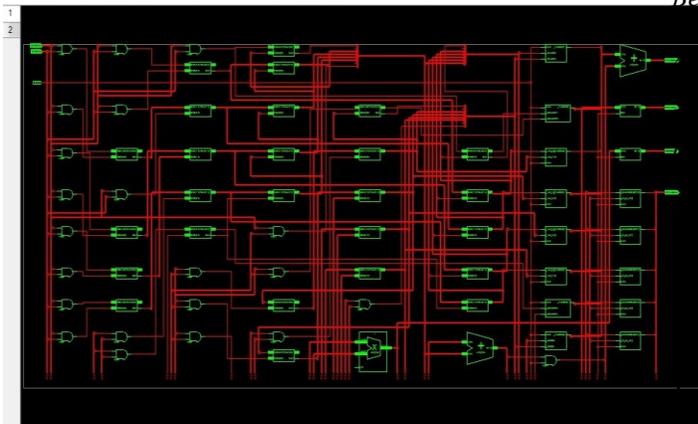

*Fig4. RTL result of reconfigurable FFT*

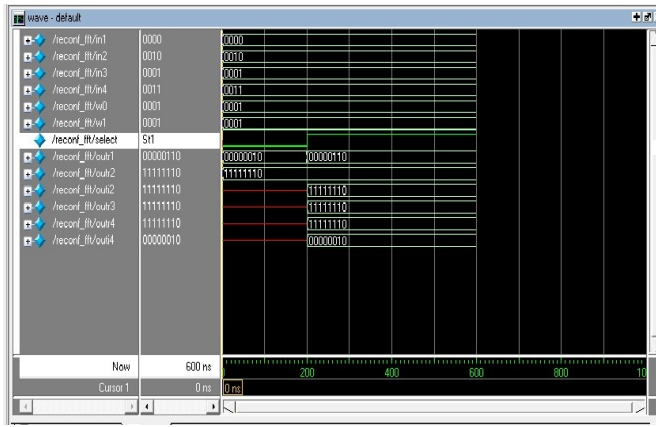

*Fig5. Simulation result of reconfigurable FFT*

## CONCLUSION:

An FFT circuit has been described that provides the high performance (throughput), dynamic range, Small area, functionality, and flexibility that can benefit future needs of wireless communications systems. In particular it provides for run-time FFT transform length selection, pruning to support OFDMA protocols, and non-power-of-two FFT sizes. This is achieved with a simple, localized, regular circuit which minimizes overall system support costs associated with design, test and maintenance.

Dr Rakesh Kumar Sarin was born on September 5, 1955 at New Delhi, India. He did his BSc (Engg) ECE, from NIT–Kurukshetra in 1978, ME (ECE) from IIT-Roorkee in 1980 and PhD from A F Ioffe Physico Technical Institute St Petersburg (Russia) in 1987. He has interests in semiconductors, optoelectronics, microelectronics /VLSI, microwaves and RF ; and has published papers in these areas. He had worked at IIT-Kharagpur and at SAMEER in India ; and at University of Sheffield in the UK. He joined Electronics and Communication Engineering (ECE) Department of NIT-Jalandhar (India) in 1994. He has been working as Head of the ECE Department since November 1994.
Dr Sarin has been associated with IEEE since 1978-79, first as a Student Member and subsequently, as a Member. At IIT-Kharagpur He was Executive Member in 1988-89 of Locall IEEE body He is an FIETE, also.

Ashish Raman was born on May 15, 1983 at Moradabad, India. He did his BE ,(ECE), from MIT–Moradabad in 2003, M.Tech (Microelectronics and VLSI Design) from SGSITS -Indore in 2005 and persuing PhD from NIT Jalandhar under the guidance of Dr Rakesh Kumar Sarin . He has interests in semiconductors, microelectronics /VLSI, RF,Embedded System,DSP ; and has published papers in these areas. He had worked at NIT -Durgapure and IMSEC Ghaziabad India. He joined Electronics and Communication Engineering (ECE) Department of NIT-Jalandhar (India) in 2007. He has been working as Assistant Professor the ECE Department. He has been associated with IACSIT since 2010

Anvesh Kumar was born on April 12, 1987 at Delhi, India. He did his BE ,(ECE), from DNP COE–Shahada in 2008, persunig M.Tech (VLSI Design) from NIT - Jalandhar under the guidance of Ashish Raman . He has interests in microelectronics /VLSI, Embedded System,DSP

,ASIC Design,Digital System Design; and has published papers in these areas